# Controlling slow and fast light and dynamic pulse-splitting with tunable optical gain in a whispering-gallery-mode microcavity

M. Asano,[1] Ş. K. Özdemir,[1,2,a)] W. Chen,[2] R. Ikuta,[1] L. Yang[2], N. Imoto[1],
T. Yamamoto[1,a)]

[1]*Department of Material Engineering Science, Graduate School of Engineering Science, Osaka University, Toyonaka,*

*Osaka 560-8531, Japan*

[2]*Department of Electrical and Systems Engineering, Washington University, St. Louis, MO 63130, USA*

**We report controllable manipulation of slow and fast light in a whispering-gallery-mode (WGM) microtoroid resonator fabricated from Erbium ($Er^{3+}$) doped silica. We observe continuous transition of the coupling between the fiber-taper waveguide and the microresonator from undercoupling to critical coupling and then to overcoupling regimes by increasing the pump power even though the spatial distance between the resonator and the waveguide was kept fixed. This, in turn, enables switching from fast to slow light and vice versa just by increasing the optical gain. An enhancement of delay of two-fold over the passive silica resonator (no optical gain) was observed in the slow light regime. Moreover, we show dynamic pulse splitting and its control in slow/fast light systems using optical gain.**

Whispering-gallery-mode (WGM) microresonators have found a wide range of applications due to their high quality factors Q (long photon storage time, low loss and narrow linewidth), microscale mode volume V (tight spatial confinement leading to resonantly enhanced light intensity) and high-finesse F (strong resonant build-up of optical power). They have been used in cavity-enhanced metrology and sensing[1-4], on-chip low threshold microlasers[5-8], optomechanics[9-12], cavity quantum electrodynamics[13,14] and parity-time symmetric photonics[15,16].

Slow and fast light phenomena were observed in a solitary microcavity[17,18], an array of directly coupled microcavities[19,20], and photonic crystals[21]. This is not surprising because slow and fast light are associated with strong dispersion and can be seen in many different physical systems which involve resonant transmission. Tunable control of slow and fast light has been implemented using nonlinear optical gain, such as Raman and Brillouin gain, to realize tunable delay lines in integrated photonic systems[22-25]. Whispering-gallery-mode (WGM) microcavities are ideal for controlling the speed of light due to their sharp resonances. It has been demonstrated that controlling the coupling strength between a microcavity and a waveguide (e.g., fiber taper) provides tunability from slow-to-fast and fast-to-slow light. The coupling between the microcavity and the waveguide can be in three different regimes: First is the undercoupling regime where the coupling strength $\kappa_{ext}$ is weaker than the intrinsic losses $\kappa_0$ of the microcavity (i.e., $\kappa_0 > \kappa_{ext}$), and a light pulse coupled to the system experiences advancement

(i.e., fast light). Second is the critical coupling regime where $\kappa_0 \cong \kappa_{\text{ext}}$ and the transmission shows a resonance dip closer to zero-transmission due to destructive interference between the light directly transmitted to the detector and the light which gains a $\pi$- phase shift after coupled into and out of the resonator. Third is the over-coupling regime where $\kappa_0 < \kappa_{\text{ext}}$, and a light pulse coupled to the system experiences delay (i.e., slow light).

In a typical experiment, one can drive the waveguide-resonator system from one coupling regime to the other and hence from slow-to-fast or fast-to-slow light regimes by tuning the distance between the waveguide and the resonator (i.e., increasing or decreasing the airgap between them)[18,19]. However, this is not a practical process. For example, in on-chip silicon photonics, the distance between the waveguide and the resonator is fixed at the time of fabrication and it cannot be changed. In fiber-taper coupled resonator systems, one can use nanopositioners to tune this distance; however, this mechanical tuning is slow and may introduce mechanical oscillations and instability. Thus, alternative methods which can enable fast and stable switching between slow and fast light are needed. The maximum attainable delay or advancement and the attenuation of optical pulses during transmission are two important issues to be considered in practical implementations. The former depends on the steepness of the dispersion curve around the resonance, and can be varied by controlling the linewidth of the resonance. The latter, on the other hand, is related with the loss of the resonators which is ultimately limited by the material absorption, and can be partly compensated by sending the light slightly off-resonance.

It is well-known that introducing optical gain into a microresonator can help to compensate some of the intrinsic losses. Microresonators with optical gain are referred to as active resonators and have been used as on-chip microlasers[6,26], finesse control[27], and high sensitivity sensors[28,29] due to their enhanced Q-factor, which is the result of linewidth narrowing due to the compensation of intrinsic losses by the optical gain. Recently it has been shown that by tuning the gain in active resonators one can controllably transit between different coupling regimes without the need for physically moving the waveguide or changing the airgap between the resonator and the waveguide mechanically[30,31]. Inspired by these works, here we report the demonstration of gain-assisted slow and fast light in a microtoroid resonator fabricated from Erbium ($Er^{+3}$) doped silica. We show controllable transition among different coupling regimes by tuning the gain which also enables switching between slow and fast light. Moreover, we report two-fold improvement in the delay of the pulses, and present results showing that the provided-gain also allows the transmission of pulses with little or no loss.

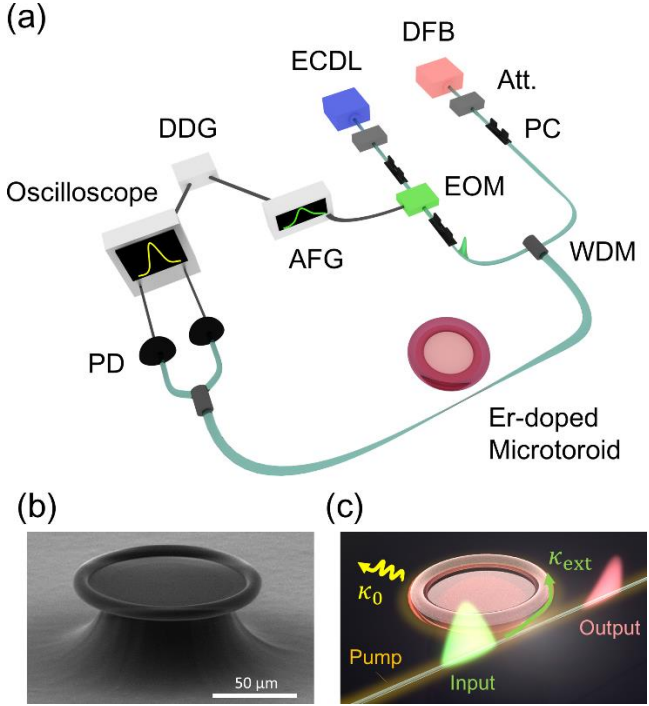

*FIG. 1. (a) Schematic illustration of the experimental setup. ECDL: External cavity diode laser; DFB: Distributed feedback laser; Att.: Attenuator; PC: Polarization controller; EOM: Electro-optic modulator; AFG: Arbitrary function generator; DDG: Digital delay generator; WDM: wavelength division multiplexer; PD: photodetector; Er: Erbium. (b) Scanning electron microscope image of the microtoroid used in the experiments. (c) Illustration of our experiment observing slow and fast light of optical pulses. $\kappa_0$: intrinsic loss of the resonator (i.e., includes radiation, scattering and material absorption losses), and $\kappa_{ext}$: cavity-waveguide coupling loss. Pump provides the optical gain by exciting Er ions to compensate losses. Input light is either delayed ($\kappa_0 < \kappa_{ext}$) or advanced ($\kappa_0 > \kappa_{ext}$).*

The setup used in our experiments is shown in Fig. 1a. The active resonator was a microtoroid resonator fabricated from $Er^{3+}$ doped silica. The major and minor diameters of the microtoroid was about 100 μm and 7 μm, respectively. The height of the silicon pillar was around 60 μm. A fiber taper fabricated by the heat-and-pull technique[32] with a subwavelength diameter was used to guide laser light into and out of the WGM of the microtoroid. Figure 1b show the scanning electron microscope image of the microtoroid used in the experiments. The pump laser used to excite the $Er^{3+}$ ions was a DFB laser in the 1480 nm band. When pumped in this band, $Er^{3+}$ ions provided gain photons compensating the losses of the resonances in the 1550 nm band. Fig. 1c is an illustration of the experiment. Using an external cavity tunable laser in the 1550 nm band with a linewidth of 300 kHz as the probe, we have confirmed that as the pump power was increased (i.e., optical gain was increased) the linewidths of the resonances in the 1550 nm band became narrower and Q of the mode increased (Fig. 2a). As the pump power was increased, the resonance linewidth monotonously decreased until it was no longer affected by the gain (Fig.2b), indicating gain saturation. An interesting observation regarding Fig. 2a is that as the power was increased, the dip of the resonance first approached zero (i.e., no or little transmission at resonance) and then started moving away from zero (i.e., non-zero transmission at the resonance). This implies that the gain shifted the resonator-taper system from the undercoupling towards the critical coupling and then to the overcoupling regime. This can be understood as follows. In the undercoupling regime we have $\kappa_0 > \kappa_{ext}$ indicating higher intrinsic losses than coupling losses. Since the distance between the taper and the resonator was kept fixed, the coupling losses $\kappa_{ext}$ did not change, whereas the intrinsic loss $\kappa_0$ decreased with increasing gain, thereby approaching $\kappa_{ext}$. This shifted the system from undercoupling to close to critical coupling where $\kappa_0 \cong \kappa_{ext}$. This is reflected in the transmission spectra as a transition from a higher transmission at resonance to close-to-zero transmission. Further increase of the gain (i.e., decrease of $\kappa_0$) pushed the system into the overcoupling regime gradually as $\kappa_0$ became smaller and smaller, satisfying $\kappa_0 < \kappa_{ext}$. As a result, the transmission at resonance increased. We note that the Fano-like asymmetric lineshape observed in the overcoupling regime (red curve, Fig. 2a) is due to the time-dependent gain profile (i.e., dynamic gain excitation) induced by sweeping the wavelength of the pump laser across the resonance[33].

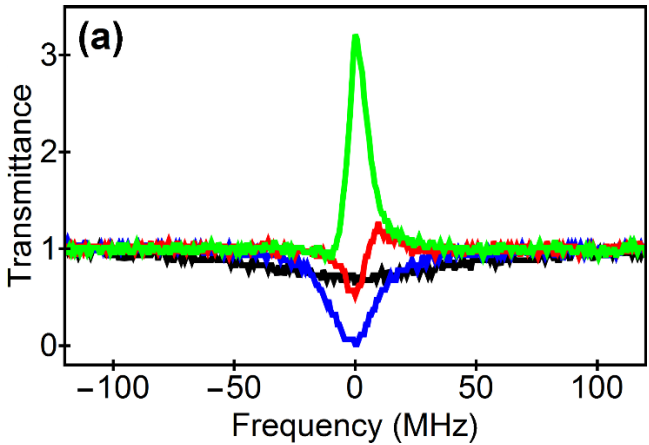

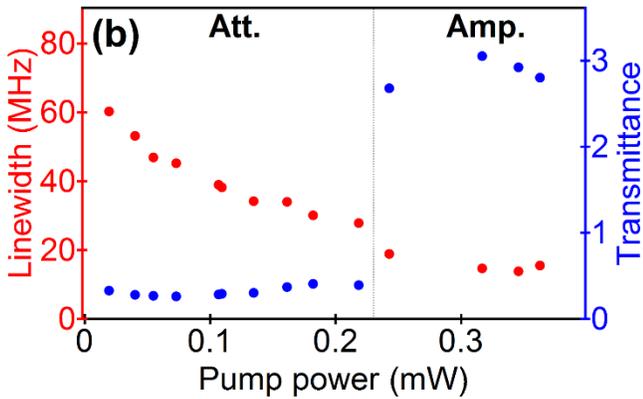

*FIG. 2. Effect of the optical pump power (i.e., optical gain) on the fiber-taper coupled Erbium-doped silica microtoroid. (a) With increasing gain (pump power), the system initially in the undercoupling regime (black curve) moved towards critical coupling (blue curve) and then to overcoupling (red curve) regime. Further increase of the power led to amplification with a pronounced resonance peak (green curve). (b) Measured linewidth and the transmission at the resonance dip (or the resonance peak) as a function of pump power.*

After confirming that the gain helps to move the system into different coupling regimes, we performed experiments to check how this would affect the transmission of light pulse. The light pulses were prepared by amplitude modulation of the cw probe laser using an electro-optical modulator (EOM). First, a reference measurement was done using the fiber-taper waveguide without the resonator (the distance between the resonator and the fiber-taper was so large that there was no coupling between them). The pulse at the output of the fiber-taper was detected by a PD connected to an oscilloscope. Next, keeping all the settings the same, we brought the resonator closer to the fiber-taper so that they can exchange energy. The light out-coupled from the resonator was sent to the PD. The oscilloscope recorded the detected signal in comparison to the reference signal. The difference in the time-of-arrivals of the pulse-peaks of the reference and the signal indicated delay or advancement induced by the resonator.

We performed measurements by varying the pump power (i.e., optical gain) for various initial coupling conditions (Figs. 3a and 3b). In these measurements, we used thermal locking to stabilize the frequency of the pump laser in a resonance. The pump power was kept below the lasing threshold of the active resonator. At each pump power, we finely tuned the frequency of the probe pulses to obtain maximum temporal shift. At the same time we measured the peak intensity of the transmitted pulses and normalized it with the peak intensity of the input pulses to assign an amplification factor: unity amplification factor implies lossless transmission; an amplification factor less than one implies loss, and finally an amplification factor larger than one implies that the optical gain provided by the pump overcomes all the losses in the system and amplifies the transmitted pulses. Figure 3 presents typical results depicting the evolution of the coupling regimes and the measured group delay as a function of the pump power for a pulse with a width of 12.5 ns. The transition from undercoupling to overcoupling and then into the amplification regime is clearly seen with increasing optical gain. As seen, when the pump power was low, the systems was in the undercoupling regime and the light pulses going through the resonator experienced advancement, indicated by negative group delay in Figs. 3a and 3b. When the system was moved into the overcoupling regime by increasing gain, pulses experienced delay (slow light) as indicated by positive group delay. The group delay stayed constant when the pump power reached a critical value (Fig. 3b). We attributed this to gain saturation. The effect of the initial coupling distance between the resonator and the fiber-taper reflected itself as the required pump power (or gain) to shift the system from fast to slow light and vice versa. It also determined the amount of loss compensation and amplification as the optical pulse propagates through the system. As seen in Figs. 3a and 3b, peak pulse intensity decreased as the system was moved from undercoupling regime to critical coupling with increasing pump power. When the power was further increased and the system completed its transition into the overcoupling regime, peak pulse intensity increased. This is in good agreement with the loading curve[34] (transmission versus waveguide-resonator coupling distance) of waveguide-coupled microresonators where transmission decreases from one to zero as the distance between the waveguide and the resonator is decreased to approach critical coupling where transmission becomes zero. If the distance is decreased further, the system moves into overcoupling regime and transmission increases close to one. In our system, the presence of gain allows us to overcome the losses and amplify the signal when the system is in the overcoupling regime.

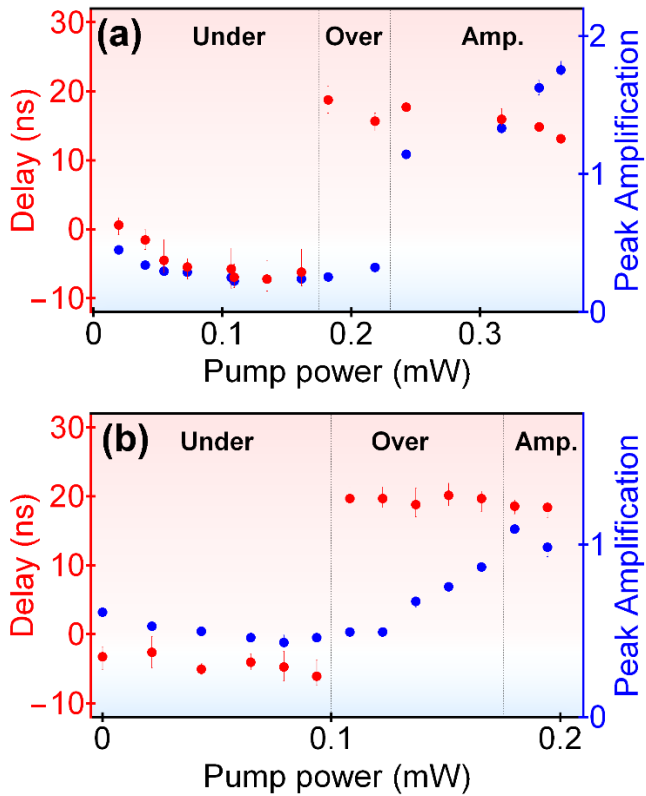

FIG. 3. *Effect of the optical pump power (i.e., optical gain) on the delay (red) and the peak pulse intensity (blue) for 12.5 ns optical pulses coupled into the Erbium-doped silica microtoroid resonator. (a) and (b) differs only in the initial coupling condition. Negative value for delay implies the advancement of the pulses.*

In Fig. 4a we present a series of time-domain pulse profiles showing the evolution of the pulse shape as the pump power (i.e., optical gain) is increased in an active (i.e., with optical gain from Erbium ions) microtoroid which was initially set in the undercoupling regime. When the pump laser was off, the pulse experienced advancement (i.e., negative delay) because the system was set to undercoupling. Increasing the pump power shifted the system closer to the critical coupling regime resulting in more advancement of the pulse. This continued until the pump power reached at *P=0.11* mW, when dynamic pulse splitting was observed, implying that the system was set at critical coupling. This pulse splitting stems from the difference in the dynamic behavior of the light ballistically transmitted through the waveguide (without coupling into the resonator) and the light coupled into the waveguide after circulating in the resonator[18,35]. At the critical coupling condition, the portions of the ballistic and the circulated pulses that overlap temporally cancel each other; however, the trailing edge of the circulated pulse and the leading edge of the ballistic pulse do not overlap temporally and hence cannot cancel each other, leading to the observed splitting. One of the split pulses exhibits advancement and the other one exhibits delay with respect to the pulse propagation in the absence of the resonator. If the pulse has an ideal Gaussian shape and the system is at ideal critical coupling, the advanced and delayed pulses will have the same shapes and pulse peaks (i.e., balanced splitting). Deviations from ideality then results in an unbalanced splitting similar to what we observed in our experiments. It should be noted that the splitting-balance at critical coupling depends on the temporal width of the optical pulses[18]. In our experiments, we chose the temporal widths of the pulses so that delay is maximized. For such pulses, splitting is not balanced as seen in Fig. 4.

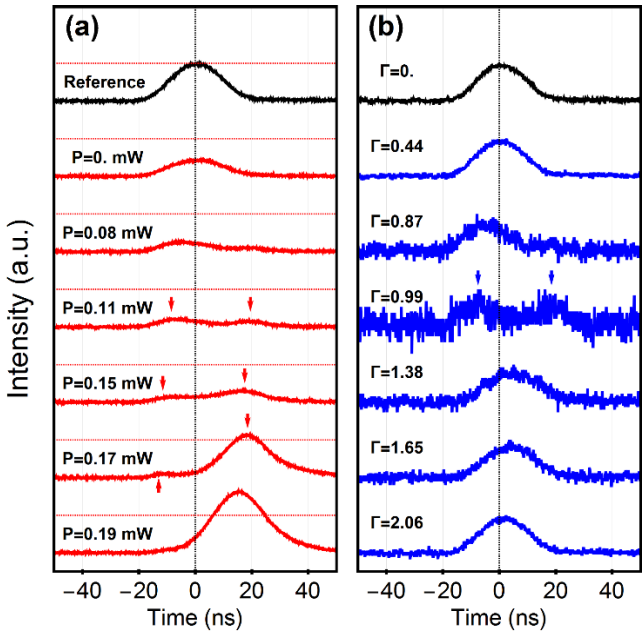

*FIG. 4. Dynamical pulse splitting and its control via (a) optical gain, and (b) mechanical positioning Evolution of the shape of 12.5 ns optical pulses as they were transmitted through a fiber taper-coupled is shown in (a) for an active resonator as the pump power (optical gain) was varied, and in (b) for a passive resonator as the coupling condition was varied. The intensity of pulses in (b) are normalized by their peak value for clarifying the pulse shapes. The black dotted vertical lines show the peak position of the reference pulse and the red dotted horizontal lines in (a) show the peak intensity of the reference pulse. Arrows indicate the split pulses. Note the loss compensation observed in the active resonator at P=0.17 mW and amplification at P=0.19 mW in (a). In (b) Γ is the coupling parameter measured at each coupling distance between the resonator and the fiber taper.*

In our experiments, dynamic pulse splitting was unbalanced to favor advancement (i.e., advanced pulse had higher pulse peak) at low pump power because the system was close to the critical coupling but at the undercoupling side. As the pump power was increased and the system was pushed through the critical coupling point into the overcoupling side, the splitting favored the delayed pulse (i.e., delayed pulse had higher peak. See the spectrum for *P=0.15* mW in Fig. 4a). As the system was pushed gradually away from the critical coupling point into the overcoupling regime, the splitting was minimized and faded away completely at *P=0.19* mW, resulting in a delayed pulse. This observation suggests the use of optical gain for controlling and tuning dynamic pulse splitting in slow and fast light systems. Additionally, the existence of the gain helps to compensate the losses that the optical pulses experience when transmitted in slow/fast light systems, as evidenced with the amplification of the delayed pulse in our experiments.

In Fig. 4b, we present a series of pulse shape spectra which were obtained for a passive (i.e., no gain) microtoroid resonator for various values of the coupling parameter $\Gamma \equiv Q_{\text{intrinsic}} Q_{\text{loaded}}^{-1} - 1$. Here $Q_{\text{intrinsic}}$ is the intrinsic quality factor (taking into account all the losses in the resonator except for the coupling loss) of the resonance and was measured by setting the taper-resonator system at deep undercoupling regime where the coupling losses are negligible. $Q_{\text{loaded}}$ is the loaded quality factor of the same resonance, which takes into account both the intrinsic and the coupling quality factors, and it was calculated from the measured linewidth of the resonance at each coupling distance between the resonator and the taper. Dynamic pulse splitting, similar to what was observed in an active resonator (Fig. 4a), was seen for the passive microtoroid as we mechanically, using a piezo-stage, changed the coupling condition from undercoupling regime to critical coupling and then to overcoupling regime. This implies that slow and fast light as well as dynamical pulse splitting are strongly dependent on the coupling regime and how far the coupling condition is away from the critical coupling point. This also provides another evidence that coupling condition and the related physical processes in a waveguide-coupled resonator system can be controlled just by tuning optical gain.

In conclusion, we have presented experimental results showing that gain provided by dopants in a WGM microresonator can effectively compensate for material losses enabling high-Q optical modes and controllable waveguide-resonator coupling for selective transition from fast to slow light or vice versa. Optical gain and its modulation also provide a mechanism to control dynamic pulse splitting and distortions that optical pulses experience when transmitted in a slow/fast light system. We note that in the undercoupling regime we have a fast-light system and the optical gain does not help to compensate losses. However, as the gain increases so that the system moves into the overcoupling regime, we have slow-light, and the optical gain helps not only to compensate losses and amplify the transmitted light pulses but also to tune the amount of delay. Thus, the gain in the overcoupling regime enables implementing tunable delay lines. In practical applications one should consider the speed at which the system can be tuned between different regimes, and the bandwidth of operation. In our system, the speed of switching between different regimes depends on how fast we can switch the optical gain and hence the pump power. With the use of the state-of-the-art electro-optic amplitude modulators (with GHz bandwidths), the pump power can be switched between its two values at GHz rates. Bandwidth of operation, on the other hand, is limited by the linewidth of the resonance, which also affects the amount of delay (i.e., slow light): The narrower is the linewidth, the larger is the delay and the narrower is the operation bandwidth. This is a problem not only for the resonator-based slow-light systems but also for the systems based on nonlinear optical gain[22-25] (i.e. Brillouin and Raman scattering) which has limited gain bandwidth. The presence of gain in our system allows us to tune the linewidth of the resonance. Thus, depending on the needs of the application or the problem, we can trade delay with bandwidth or bandwidth with delay by adjusting the amount of optical gain. Without the optical gain, the resonance linewidth will be set at the time of fabrication by the material absorption, scattering, radiation and coupling losses with no means to adjust or change.

[a] Authors to whom correspondence should be addressed. Electronic mail: ozdemir@wustl.edu; yamamoto@mp.es.osaka-u.ac.jp.

**ACKNOWLEDGMENTS** This work was supported by MEXT/JSPS KAKENHI Grant Number 16H02214, 15H03704, 15KK0164, 16H01054. L.Y. and Ş.K.Ö. are supported by ARO grant No. W911NF-12-1-0026.